\documentclass[12pt,preprint]{aastex}

\usepackage{booktabs}
\newcommand{\HI}{\ion{H}{1}}
\newcommand{\HII}{\ion{H}{2}}
\newcommand{\Htwo}{H$_{2}$}
\newcommand{\LCO}{L$_{\rm CO}$}
\newcommand{\MHtwo}{M$_{\rm H_{2}}$}
\newcommand{\kms}{km s$^{-1}$}

\begin{document}
\title{CARMA CO Observations of Three Extremely Metal-Poor, Star-Forming Galaxies}

\author{Steven R. Warren}
\affil{Department of Astronomy, University of Maryland, College Park, MD 20742, USA; 
swarren@astro.umd.edu}

\author{Edward Molter}
\affil{Department of Physics \& Astronomy, Macalester College, 1600 Grand Avenue, Saint Paul, MN 55105, USA;
emolter@macalester.edu}

\author{John M. Cannon}
\affil{Department of Physics \& Astronomy, Macalester College, 1600 Grand Avenue, Saint Paul, MN 55105, USA;
jcannon@macalester.edu}

\author{Alberto D. Bolatto}
\affil{Department of Astronomy, University of Maryland, College Park, MD 20742, USA; 
bolatto@astro.umd.edu}

\author{Elizabeth A. K. Adams}
\affil{Netherlands Institute for Radio Astronomy (ASTRON), Postbus 2, 7990 AA, Dwingeloo, The Netherlands;
adams@astron.nl}

\author{Elijah Z. Bernstein-Cooper}
\affil{Department of Astronomy, University of Wisconsin, 475 N Charter Street, Madison, WI 53706, USA;
ezbc@astro.wisc.edu}

\author{Riccardo Giovanelli}
\affil{Center for Radiophysics and Space Research, Space Sciences Building, Cornell University,
Ithaca, NY 14853, USA;
riccardo@astro.cornell.edu}

\author{Martha P. Haynes}
\affil{Center for Radiophysics and Space Research, Space Sciences Building, Cornell University,
Ithaca, NY 14853, USA;
haynes@astro.cornell.edu}

\author{Rodrigo Herrera-Camus}
\affil{Department of Astronomy, University of Maryland, College Park, MD 20742, USA;
rhc@astro.umd.edu}

\author{Katie Jameson}
\affil{Department of Astronomy, University of Maryland, College Park, MD 20742, USA;
kjameson@astro.umd.edu}

\author{Kristen B. W. McQuinn}
\affil{Minnesota Institute for Astrophysics, University of Minnesota, Minneapolis, MN 55455,
USA;
kmcquinn@astro.umn.edu}
\affil{McDonald Observatory; The University of Texas; Austin, TX 78712, USA}

\author{Katherine L. Rhode}
\affil{Department of Astronomy, Indiana University, Bloomington, IN 47405, USA;
rhode@astro.indiana.edu}

\author{John J. Salzer}
\affil{Department of Astronomy, Indiana University, Bloomington, IN 47405, USA;
slaz@astro.indiana.edu}

\author{Evan D. Skillman}
\affil{Minnesota Institute for Astrophysics, University of Minnesota, 116 Church St. SE, Minneapolis, MN 55455; 
skillman@astro.umn.edu}

\begin{abstract}

We present sensitive CO (J = 1 $\rightarrow$ 0) emission line
observations of three metal-poor dwarf irregular galaxies Leo~P
($Z$$\sim$3\% $Z_{\odot}$), Sextans~A ($Z$$\sim$7.5\% $Z_{\odot}$),
and Sextans~B ($Z$$\sim$7.5\% $Z_{\odot}$), all obtained with the
Combined Array for Millimeter-wave Astronomy (CARMA) interferometer.
While no CO emission was detected, the proximity of the three systems
allows us to place very stringent (4$\sigma$) upper limits on the CO
luminosity (\LCO) in these metal-poor galaxies. We find the CO
luminosities to be \LCO\ $<$ 2900 K km/s pc$^2$ for Leo~P, \LCO\ $<$
12400 K \kms\ pc$^2$ for Sextans~A, and \LCO\ $<$ 9700 K \kms\ pc$^2$
for Sextans~B. Comparison of our results with recent observational
estimates of the factor for converting between \LCO\ and the mass of
molecular hydrogen, as well as theoretical models, provides further
evidence that either the CO-to-\Htwo\ conversion factor increases
sharply as metallicity decreases, or that stars are forming in these
three galaxies very efficiently, requiring little molecular hydrogen.

\end{abstract}  

\section{Introduction \label{intro}}

Metal-poor environments such as those in low-mass dwarf galaxies and
the outskirts of normal spiral galaxies are chemically similar to the
star-forming environments of the early universe. Improving our
understanding of these environments will help shed light on some of
the processes that drive star formation in the early universe.  It is
clear that heavy elements help cool interstellar gas to initiate its
eventual collapse into potential star-forming regions (e.g.,
\citealt{spit48}; \citealt{shu87}; \citealt{mcke89}; \citealt{wolf95};
\citealt{glov07}).  How does star formation proceed without the
presence of heavy elements?  Does star formation require large
concentrations of molecular hydrogen (\Htwo), or is this molecule just
a byproduct of the star formation process at low metallicity?

In a solar metallicity environment like that of the Milky Way,
molecular hydrogen easily forms on the available dust grains. Directly
observing the \Htwo\ gas that will form stars is problematic since
\Htwo\ does not radiate at temperatures below a few hundred Kelvin,
while stars form inside cold molecular clouds with temperatures of a
few tens of Kelvin. Fortunately, CO forms in conditions similar to
\Htwo, and the luminosity of CO (\LCO) is correlated to the total mass
of \Htwo\ (\MHtwo; \citealt{youn82}; also see the recent review by
\citealt{bola13} and references within).

While the conversion factor between \LCO\ and \MHtwo\ is fairly well
established in normal galaxy disks with metallicity approximately
equal to the solar value, it is much less certain at low metallicities
(\citealt{malo88}; \citealt{isra97}; \citealt{lero11};
\citealt{bola13}). At low metallicity, dust is much less abundant
(e.g., \citealt{lise98}; \citealt{gala12}).  The dust on which
\Htwo\ forms also plays a critical role in the shielding of the
molecular gas from photodissociation by UV photons. The lower dust
abundances also lengthen the time for \Htwo\ to reach chemical
equilibrium by up to 1 Gyr \citep{bell06,glov11}. This raises the
question of whether atomic gas clouds can form stars in the absence of
a significant molecular component.  It is likely that for modest
changes in metallicity from a solar abundance this is not a problem
since [\ion{C}{2}] can provide most of the needed cooling
\citep{krum12,krum13}, and during collapse, the increase in density
drives much of the gas into the molecular phase \citep{glov12b}.

There has been extensive effort aimed at both observing and modelling
these low metallicity environments. \citet{tayl98} observed the CO
emission in 11 nearby, low metallicity dwarf galaxies and found a
striking dropoff of detections around an oxygen abundance of 12 +
log(O/H) $\lesssim$ 8.0. This same limit was seen by \citet{schr12} in
16 dwarf galaxies from the HERACLES survey \citep{lero09}.  The lack
of CO detections at 12 + log(O/H) $\lesssim$ 8.0 could indicate the
complete absence of CO or it could point to our limited technological
ability to recover the low surface brightness emission. The recent
detection of CO emission in WLM (12 + log(O/H) = 7.8) by
\citet{elme13} and marginal detection of CO in Sextans~A by
\citet{shi15} both using the Atacama Pathfinder EXperiment (APEX;
\citealt{gust06}) suggests the latter. Nevertheless, it is clear that
the abundance of CO drops as the metallicity decreases.

Simulations of star formation at low metallicity have shown that
molecular hydrogen may form and become abundant prior to any CO
formation, even forming in gas with no dust or heavy elements
(\citealt{glov12a}; \citealt{glov12b}; see also, for example,
\citealt{krum08}; \citealt{krum09a}; \citealt{krum09b};
\citealt{ostr10}; \citealt{krum12}; \citealt{krum13}).  We aim to
place further constraints on the state of the molecular interstellar
medium (ISM) at low metallicity in relation to the star formation
activity by observing the CO (J = 1 $\rightarrow$ 0) emission line in
three nearby, low-metallicity, star-forming dwarf galaxies. We
describe our observations in \S2 and our galaxy sample in \S3.  We
discuss our results in \S4, compare our galaxy sample to those in the
literature in \S5, and summarize our findings in \S6.

\section{Observations and Data Processing \label{data}}

We observed the CO (J = 1 $\rightarrow$ 0) 115.27120 GHz emission line
in three extremely metal-poor galaxies (Leo~P, Sextans~A, and
Sextans~B) with the Combined Array for Millimeter-wave Astronomy
(CARMA) in the D-configuration. Because the CARMA primary beam
($\sim$1 arcmin) covered the majority of the star forming disk in Leo
P, it was observed with a single pointing for a total of 24 hours
between 11 February 2013 and 16 February 2013.  A seven-pointing
mosaic was used to observe portions of Sextans~A (17 total hours
between 2-15 June 2013) and Sextans~B (11 total hours between 28 March
2013 and 2 April 2013) that contained recent star formation activity
as well as dense, potentially cold \HI\ gas \citep{warr12}. Figure 1
shows optical images of each galaxy. The contours represent the H I
column density and the red circles denote the field of view of our
CARMA observations in each galaxy.

The correlator was set up to provide native channel widths of 2
\kms\ (0.78 MHz) for each galaxy. Standard reduction procedures were
followed using the Miriad software package.  Baseline and rest
frequency corrections were applied, high-amplitude data were flagged,
and the sources were calibrated against flux (3C273) and phase
(0854+201, 1058+015) calibrators. The final data cubes have a natural
weighted beam size of 3.5\arcsec$\times$2.5\arcsec and a velocity
range of 600 \kms\ centered near the velocity at which \HI\ emission
was detected for each galaxy (\citealt{ott12};
\citealt{bern14}). Table \ref{galprops} lists relevant galaxy
properties.

In addition to each of the above spectral line observations, we placed
the 12 remaining CARMA spectral windows to observe the continuum
emission. Each of these spectral win- dows covers a 500 MHz bandwidth
(6 GHz total). Similar reduction procedures as above were performed on
the continuum data. The entire 6 GHz bandwidth was then combined into
a single image for later analysis.

We complement our CO analysis with Jansky Very Large Array
\HI\ emission line observations from two other studies: \citet{bern14}
(Leo~P) and \citet{ott12} (Sextans~A and Sextans~B).  We refer the
reader to these manuscripts for full data reduction and processing
steps.  We utilize these natural weighted data sets to guide our
search for CO emission as we expect the CO emission to overlap in
frequency with the observed \HI.  In \S\ref{h2mass} we use the
\HI\ data to estimate the amount of \Htwo\ for comparison to that
computed from our CO observations.

\section{Galaxy Sample \label{galsamp}}

We observed three nearby galaxies that show recent signs of star
formation, suggesting that some amount of molecular gas is likely to
be present.  Leo~P is a recently discovered, extremely low-mass galaxy
identified by the Arecibo Legacy Fast ALFA (ALFALFA) survey
\citep{giov13}. Initial estimates for Leo~P placed it at a distance
between 1.5 and 2.0 Mpc \citep{rhod13}, while deep Large Binocular
Telescope imaging refined the distance determination to 1.62 Mpc
\citep{mcqu15}. At this distance, Leo~P has a total \HI\ mass of only
8.1$\times$10$^{5}$ M$_{\sun}$ \citep{mcqu15}.  Despite its low gas
content, Leo~P is currently forming stars as evidenced by a young,
blue stellar population as well as a single bright \HII\ region
\citep{rhod13}. There also exists evidence of a cold
\HI\ (T$\lesssim$1000 K) gas phase near this lone \HII\ region
\citep{bern14}.  Optical spectroscopic observations of the bright
\HII\ region put a firm constraint on the oxygen abundance (12 +
log(O/H) = 7.17$\pm$0.04; \citealt{skil13}), establishing Leo~P as one
of the lowest metallicity gas-rich galaxies ever measured.  Since
  Leo~P was discovered recently, no Spitzer or Herschel observations
  exist.

At distances of $\sim$1.4 Mpc \citep{dalc09}, Sextans~A and Sextans~B
are both well-studied systems.  Each has a metallicity of 12 +
log(O/H) $\approx$ 7.55 \citep{knia05} and high column density
\HI\ reservoirs (\citealt{ott12}). Their \HI\ disks show evidence of
cold \HI\ gas (T $\lesssim$ 1500 K; \citealt{warr12}), as well as
co-spatial dust emission from Spitzer imaging
\citep{dale09}. Following the methods used in {Herrera-Camus et
  al. (2012}\nocite{herr12}; specifically, a modified blackbody model
with an emissivity index $\beta$=1.5 and a mass absorption coefficient
of $\kappa$$_{\rm 250 \mu\,m}$ = 9.5 cm$^{2}$ g$^{-1}$), the
\citet{dale09} Spitzer fluxes imply dust masses of
$\sim$1.1\,$\times$10$^{3}$ M$_{\sun}$ and $\sim$240 M$_{\sun}$ for
Sextans\,A and Sextans\,B, respectively.  This dust mass for Sextans A
can be compared to the $\sim$770 M$_{\odot}$ derived from Herschel
observations in \citet{shi15}.

\begin{figure}
\includegraphics[width=155mm]{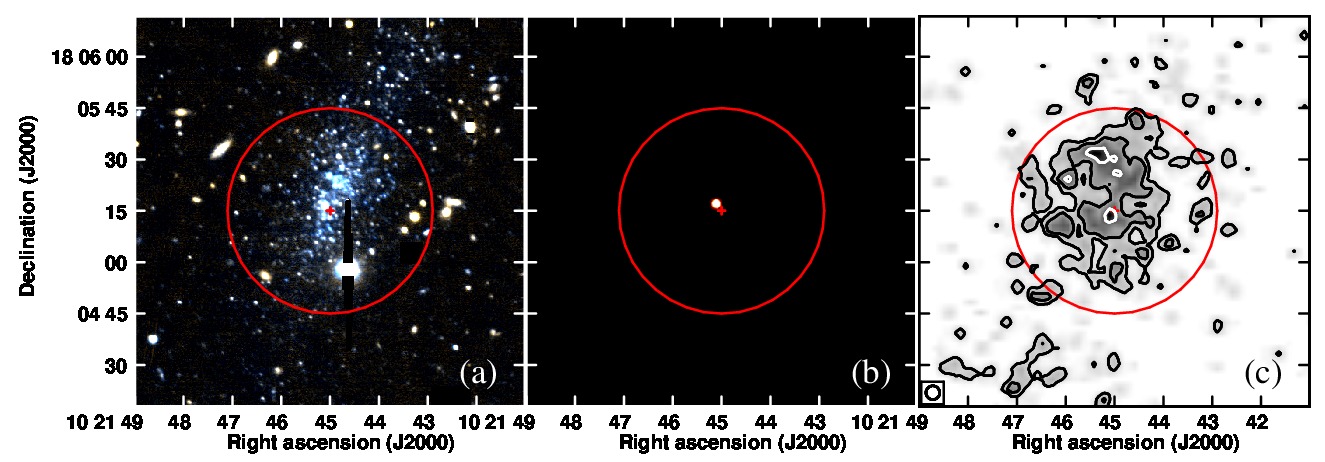}\\
\includegraphics[width=155mm]{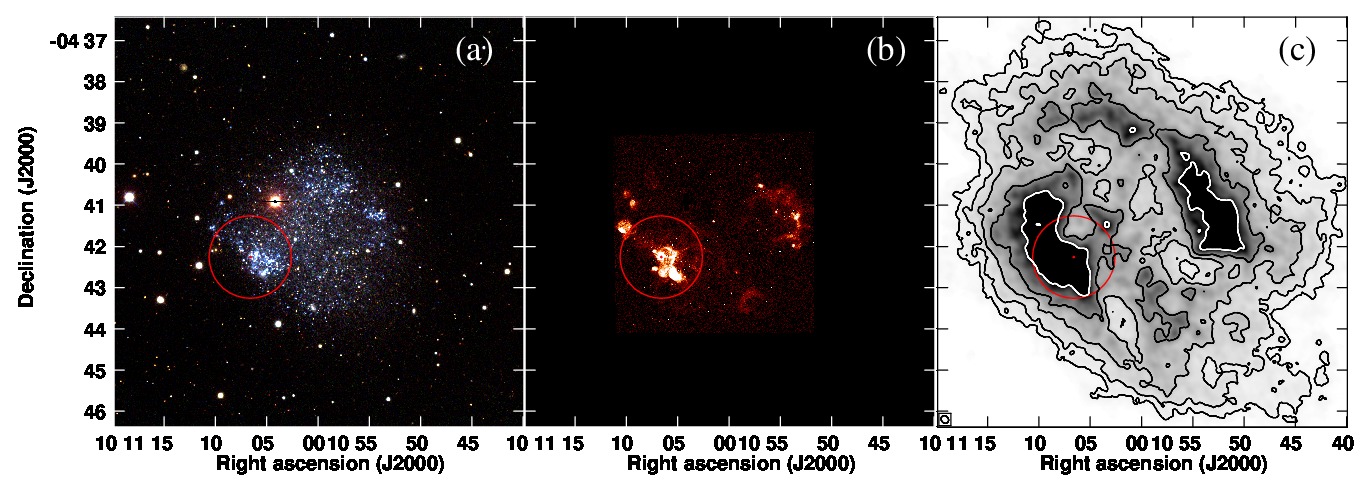}\\
\includegraphics[width=155mm]{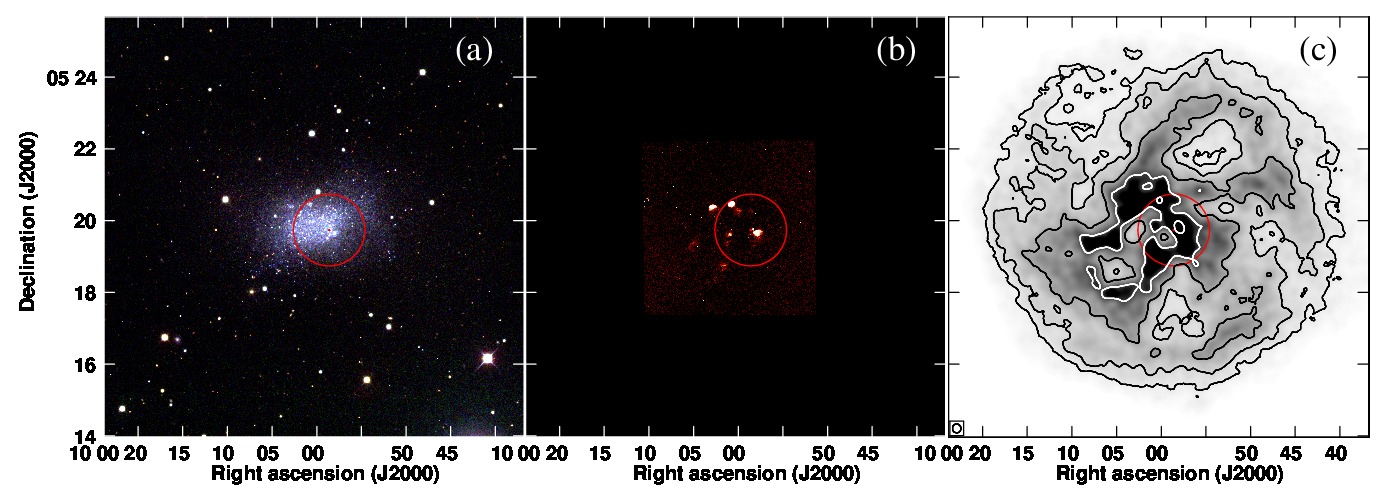}
\caption{Optical (left column), H$\alpha$ (center column), and
  integrated \HI\ intensity maps (right column) for Leo~P (top),
  Sextans~A (middle), and Sextans~B (bottom).  Contours on the
  \HI\ maps are at the 1.25, 2.5, 5, 10, and 20$\times$20 cm$^{-2}$
  levels.  The \HI\ beams are at the lower left.  The red circle in
  each panel denotes our CARMA half-power observing field of view.
  {\it Leo~P:} (a) Large Binocular Telescope optical image from
  \citet{mcqu13}, (b) H$\alpha$ image from \citet{rhod13}, and (c)
  \HI\ image from \citet{bern14}.  {\it Sextans~A:} (a) Local Group
  Survey BVR image from \citet{mass07}, (b) Local Volume Legacy
  H$\alpha$ image from \citet{kenn08}, and (c) Jansky Very Large Array
  \HI\ map from \citet{ott12}.  {\it Sextans~B:} (a) Local Group
  Survey BVR image from \citet{mass07}, (b) Local Volume Legacy
  H$\alpha$ image from \citet{kenn08}, and (c) Jansky Very Large Array
  \HI\ map from \citet{ott12}.
\label{galfig}}
\end{figure}

Despite the \HI\ richness and ongoing star formation of our sample
galaxies, to date no CO emission has been found in Sextans B or in
Leo\,P; a weak CO detection in Sextans\,A was reported in
\citet{shi15} and is discussed further below.  In Figure \ref{galfig}
we show our half-power CARMA field-of-view (red circles) overlaid onto
optical (column a), H$\alpha$ (column b), and integrated
\HI\ intensity (column c) images for Leo~P (top row), Sextans~A
(middle row), and Sextans~B (bottom row).  Our CARMA observations
cover areas with recent star formation and high-column density
\HI\ emission.

\section{Results \label{results}}

\begin{figure}
\includegraphics[width=155mm, trim= 0mm 0mm 0mm 200mm]{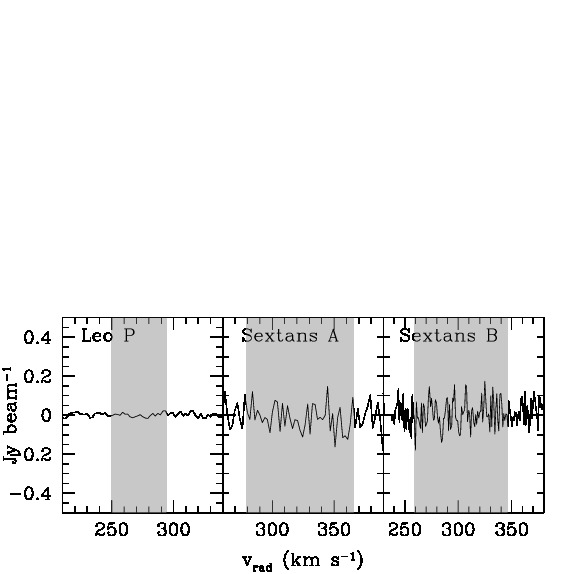}
\caption{CO spectra at 2 km s$^{-1}$ velocity resolution through the
  central 1\arcsec\ pixel of each galaxy region.  The velocity extent
  of the \HI\ spectra (see Table \ref{galprops}) has been shaded.
\label{galspec}}
\end{figure}

Visual inspection of the final data cubes at 2 \kms\ velocity
resolution (see Figure \ref{galspec}) revealed no significant CO
emission in any of the three galaxies near the expected velocity
ranges defined by the velocity ranges observed in \HI. Thus, in order
to report the upper limits of any CO emission, we follow a similar
approach as \citet{lero07}, who presented high-sensitivity
measurements of the CO emission from the extremely metal-poor galaxy I
Zw 18. First, we smooth the data to a velocity resolution of 18.4
\kms, a typical linewidth observed in other nearby, low-mass galaxies
(e.g., \citealt{schr12}). We then adopt an upper limit to the CO
intensity (S$_{\rm CO}$) of four times the average rms level in each
channel over the observed \HI\ velocity range multiplied by the
velocity resolution. To compute an upper limit to the luminosity of CO
emission (\LCO) we use the equation \LCO\ = 2453$\times$S$_{\rm
  CO}$D$_{\rm Mpc}^{2}$ where \LCO\ has units of K \kms\ pc$^{2}$,
S$_{\rm CO}$ in units of Jy beam$^{-1}$ \kms, and D$_{\rm Mpc}$ is the
distance in Mpc. Table \ref{obsprops} lists the derived upper limits
for each galaxy.

Our upper limits of the CO emission represent some of the most
sensitive to date at the respective metallicities of each
galaxy. Previous CO observations of the paradigm metal-poor galaxy
I~Zw~18 (12 + log(O/H) = 7.17; \citealt{skil93}) by \citet{lero07} and
\citet{herr12} yield an upper limit to the CO luminosity of
\LCO\ $\leq$ 10$^{5}$ K \kms\ pc$^{2}$. Our limit to the CO luminosity
in Leo~P, which has a similar metallicity to I~Zw~18 but is $\sim$10
times closer, is $\sim$30 times more sensitive.  Recently,
  \citet{shi15} observed a small region in Sextans~A (which is
  included in our observed region) and claims a marginal 3.4$\sigma$
  CO detection resulting in an L$_{\rm CO}$ lower limit of 3670 K
  \kms\ pc$^{2}$.  We discuss the implications of these limits below.

The 3 mm continuum emission in normal galaxies is dominated by
free-free emission from \HII\ regions (see \citealt{cond92}).  Our
observations are not sensitive enough to detect this emission in our
target regions even though H$\alpha$ emission exists in each region
(see Figure \ref{galfig}).  We report here our upper limits as 4 times
the rms level. The 3 mm continuum has an upper limit of 0.3 mJy for
Leo~P, 0.7 mJy for Sextans~A, and 1.0 mJy for Sextans~B.

\section{Discussion \label{discussion}}

\begin{figure}
\includegraphics[width=175mm]{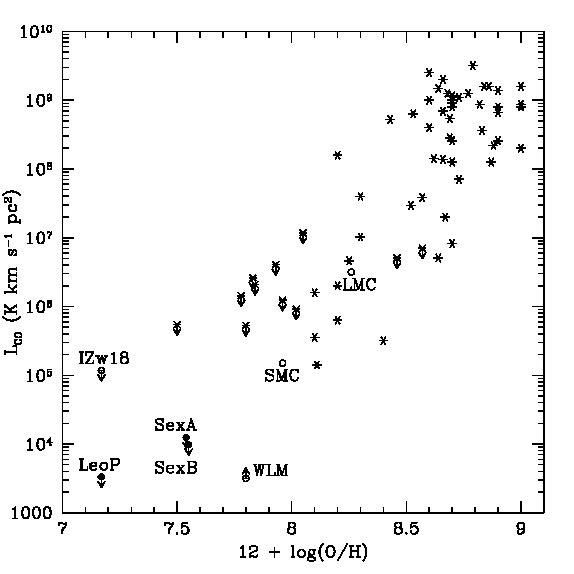}
\caption{CO (J = 1 $\rightarrow$ 0) luminosity vs oxygen abundance for
  a sample of nearby galaxies.  Black crosses are nearby galaxies
  taken from \citet{schr12} and references within (as well as the
  values for the SMC and LMC). Open circles show a few nearby galaxies
  of interest (I~Zw~18 - \citealt{lero07}; WLM -
  \citealt{elme13}). Filled black circles are the galaxies in this
  study. Detections of CO emission below metallicities of 12 +
  log(O/H) $<$ 8.0 have been mostly elusive to date.
\label{LcoMet}}
\end{figure}

It is useful to compare our CO observations to measurements of other
nearby systems in the literature.  In Figure \ref{LcoMet} we plot the
CO luminosity versus the oxygen abundance for a sample of nearby
galaxies. Black crosses are derived from the sample of galaxies
reported by \citet{schr12} and references therein. We have converted
the CO (J = 2 $\rightarrow$ 1) observations in \citet{schr12} to CO (J
= 1 $\rightarrow$ 0) by assuming the same constant (2 $\rightarrow$
1)/(1 $\rightarrow$ 0) line ratio in units of K km s$^{-1}$ of 0.7
employed by \citet{schr11} and \citet{bigi11}. Black open circles
highlight a few galaxies of interest including the I~Zw~18 limit
\citep{lero07} as well as the recently reported lower limit to the CO
emission in WLM \citep{elme13}. We correct the WLM CO (J = 3
$\rightarrow$ 2) observation to CO (J = 1 $\rightarrow$ 0) by assuming
the same (3 $\rightarrow$ 2)/(1 $\rightarrow$ 0) line ratio of 0.8 in
units of K km s$^{-1}$ employed by \citet{elme13}. Black filled
circles represent the three galaxies in this study.

It is clear from Figure \ref{LcoMet} that the detection of CO emission
below metallicities of 12 + log(O/H) $\lesssim$ 8.0 remains
difficult. If our galaxies had CO luminosities similar to the SMC we
clearly would have detected their emission.  This lack of detection
may not be due entirely to sensitivity issues, but may be the result
of changes in the physical conditions that support CO gas (e.g.,
metallicity, dust abundance, gas density, etc.).  These changes in
physical conditions have been noted several times in the literature
(e.g., \citealt{malo88}, \citealt{glov11}). The recent detection of CO
emission in the nearby galaxy WLM by \citet{elme13} suggest that newer
technology may be able to detect CO at much lower metallicities. Deep
CO observations by observatories with larger light collecting areas
such as the Atacama Large (Sub)Millimeter Array (ALMA) are needed to
establish which of the above scenarios is the limiting factor in low
metallicity CO detection.

\subsection{\Htwo\ mass estimates from CO \label{COh2mass}}

The luminosity of CO is commonly used to infer the presence of
molecular hydrogen.  To convert \LCO\ into \MHtwo\ we need to know the
relationship between the two quantities.  Substantial effort has been
expended toward understanding this correlation (see
\citealt{bola13}). For Galactic metallicities the relationship between
\LCO\ and the \Htwo\ mass, M$_{\rm H_2}$, is:
\begin{equation}\label{equ1}
{\mathrm M}_{\mathrm{H}_2} = \alpha_{\mathrm{CO}} \mathrm{L}_{\rm CO},
\end{equation}
where $\alpha_{\rm CO}$ = 4.3 M$_{\sun}$ (K \kms\ pc$^{2}$)$^{-1}$.  This conversion works 
because, at these metallicities, the CO and \Htwo\ gas are coextensive.

For galaxies with metallicities much below solar, this simple
conversion is no longer valid.  \Htwo\ can self-shield even with
little dust present, thus there should exist clouds of ``CO-faint''
\Htwo\ in star forming, low-metallicity environments which implies
that the CO and \Htwo\ are not coextensive.  Evidence for CO-faint gas
is particularly strong in the SMC \citep{rubi93, lequ94, isra97,
  bola03, lero07, lero09, lero11, bola11}.  These authors use star
formation, dust, and gas tracers to infer the presence of molecular
gas that emits only weakly in CO, if at all. The deviation away from
the Galactic value of $\alpha_{\rm CO}$ has also been inferred in CO
surveys of galaxies (e.g., \citealt{wils95, arim96, bola08, lero11,
  schr12}). Here we use our \LCO\ limits to compute upper limits to
M$_{\rm H_2}$ using a range of conversion factors.

Table \ref{h2masses} lists computed \Htwo\ mass upper limits and
the ratio of molecular-to-atomic hydrogen (M$_{\rm H_2}$/M$_{\rm HI}$)
in the observed regions for various assumptions of $\alpha_{\rm CO}$
described below. Using the Galactic value for $\alpha_{\rm CO}$
results in molecular-to-atomic hydrogen mass ratios of $\sim$1\% for
Sextans~A and Sextans~B and $\sim$5\% for Leo~P, far below those
typically observed in star-forming higher metallicity galaxies of
20-100\% (see, e.g., \citealt{lero09}). If we instead use the typical
$\alpha_{\rm CO}$ value inferred in the SMC of 70 M$_{\sun}$ (K
\kms\ pc$^{2}$)$^{-1}$ \citep{lero09}, we compute \Htwo\ masses of
$\sim$10\% of the \HI\ mass in Sextans~A and Sextans~B but $\sim$80\%
for Leo~P.  Lastly, we use the 3\% Solar metallicity models of
\citet{glov12b} to estimate M$_{\rm H_2}$.  These authors find that
the conversion factor varies in their models from $\sim$10-66.3 times
the Galactic value.  A recent ALMA study of SBS~0335-052
\citep{hunt14} suggests a lower limit value for $\alpha_{\rm CO}$
$\gtrsim$125 M$_{\sun}$ (K \kms\ pc$^{2}$)$^{-1}$ at metallicities of
$\sim$3\% solar, which falls within the range of values computed in
\citet{glov12b}.  We use the larger value from \citet{glov12b} here
which corresponds to $\alpha_{\rm CO}$ = 285 M$_{\sun}$ (K
\kms\ pc$^{2}$)$^{-1}$ to compute M$_{\rm H_2}$/M$_{\rm HI}$ ratios of
$\sim$50\% for Sextans~A and Sextans~B, and over 300\% for Leo~P.
Since low metallicity galaxies are dominated by atomic hydrogen, these
unphysical molecular-to-atomic hydrogen mass ratios either rule out
large values of $\alpha_{\rm CO}$ or, more likely, the actual CO
  emission in these galaxies is less than the derived upper limits.

\subsection{Predicting the total \Htwo\ mass from \HI\ emission \label{h2mass}}

Yet another way of predicting the amount of \Htwo\ mass in the
observed regions comes from the recent modelling of \citet{krum13}.
This model (hereafter KMT+) is an extension to the original models of
\citet{krum09b} and predicts both the \Htwo-to-\HI\ mass suface
density ratio, $f_{\rm H_2}$ = $\Sigma_{\rm H_2}$/$\Sigma_{\rm HI}$,
and the ensuing star formation rate surface density, $\Sigma_{SFR}$.
The KMT+ model assumes a two-phase ISM where chemical equilibrium may
never be reached prior to the onset of star formation in a
low-metallicity environment.  As a result, the formation of molecular
species is more of a byproduct of the collapse of star forming clouds
rather than preceded by it.  That is, the \Htwo\ is formed in the
centers of collapsing gas clouds where the temperatures are low enough
and the densities are high enough rather than the molecules forming
prior to the onset of collapse.  This scenario thus suggests that
large concentrations of molecular gas might not exist at extremely low
metallicities.

Inputs into the KMT+ model are the total gas surface density,
$\Sigma_{gas}$, the stellar + dark matter volume density
($\rho_{sd}$), the metallicity (Z), and some nominal clumping factor
(cf).  Leo~P has a stellar mass of 5.7$\times$10$^{5}$ M$_{\odot}$ and
a radius of $\sim$ 580 pc \citep{mcqu13}.  If we assume the stars have
a scale height of 100 pc then the stellar mass volume density is
approximately 0.005 M$_{\odot}$ pc$^{-3}$.  This value will change
based upon our geometric assumptions and assumptions of the dark
matter contribution. Therefore we assume three different values for
$\rho_{sd}$ in order to bracket plausible values: the value in the
Solar neighborhood $\rho_{sd}$ = 0.01 M$_{\odot}$ pc$^{-3}$
\citep{holm00} and also values an order of magnitude above and below.
\citet{krum13} suggests that cf = 1 for linear scales below $\sim$100
pc.  The \HI\ linear beam sizes are $\sim$33 pc for Leo~P, $\sim$80 pc
for Sextans~A, and $\sim$100 pc for Sextans~B, thus we assume cf = 1.
We use the output $f_{\rm H_2}$ values in combination with the
\HI\ images to estimate the total \Htwo\ masses in each region.  Table
\ref{krumh2masses} shows the total \Htwo\ mass derived for each region
in our galaxies.

We can use these mass estimates to predict what values of L$_{\rm CO}$
are expected for various assumptions of $\alpha_{\rm CO}$.  We use the
\Htwo\ mass estimates from the $\rho_{sd}$ = 0.01 models for this
exercise.  The quoted errors reflect the values for the $\rho_{sd}$ =
0.1 and 0.001 \Htwo\ mass estimates.  Table \ref{lcopreds} lists
various L$_{\rm CO}$ predictions for our regions using the same
$\alpha_{\rm CO}$ values from Table \ref{h2masses}.  If the KMT+
models are correct it seems very unlikely that CO will be observed
with current technology in Leo~P, even with ALMA.  The recent
  3.4$\sigma$ CO detection (L$_{\rm CO}$ $>$ 3670 K \kms\ pc$^{2}$)
  reported by \citet{shi15} implies an $\alpha_{\rm CO}$ $\approx$ 60
  M$_{\sun}$ (K \kms\ pc$^{2}$)$^{-1}$ in the KMT+ models.  This
  appears to be inconsistent with what is observed in objects of
  slightly higher metallicity (e.g., SMC) which require higher values
  of $\alpha_{\rm CO}$.  There is hope, however, to observe CO in
Sextans~A and Sextans~B with modest ALMA observations.

\subsection{Comparison with Star Formation Rates}

We use the SFRs computed from the H$\alpha$ and FUV luminosities to
compute limits to the amount of available molecular gas.  To do this
we need to multiply the SFR in units of M$_{\odot}$ yr$^{-1}$ by an
appropriate time scale.  \citet{weis11} detail the SFRs of both
Sextans~A and Sextans~B and show that the SFRs have remained
relatively constant over the lifetimes of each galaxy with a relative
uptick in SFR in the past $\sim$5 Gyr.  There is currently no similar
information available for Leo~P.  If we assume the SFRs have not
changed significantly in the past couple of Gyr, then we can use an
estimation of the depletion timescale, that is, the amount of time it
would take the current SFR to convert all of the available molecular
gas into stars.  \citet{bola11} find that for the SMC, the molecular
gas depletion time, $\tau_{dep}$, ranges from 0.6 - 7.5 Gyr with large
uncertainties depending on the linear scales probed.  These authors'
results are similar to the results of other studies of molecular
dominated regions in high metallicity galaxies conducted at 750 pc - 1
kpc scales which obtain $\tau_{dep}$ $\sim$ 2 Gyr
\citep{bigi08,bigi11,lero08}.  We will adopt an order of magnitude
approach to our calculation that will give us an upper limit to the
amount of available molecular gas given our assumptions of a constant
SFR over the depletion time of $\tau_{dep}$ = 2 Gyr.

With a depletion timescale of 2 Gyr and the SFR$_{\rm H\alpha}^{Reg}$
values given in Table \ref{galprops}, we compute M$_{\rm H_2}$ =
1$\times$10$^{5}$ M$_{\odot}$, 3.4$\times$10$^{6}$ M$_{\odot}$, and
6.2$\times$10$^{5}$ M$_{\odot}$ for Leo~P, Sextans~A, and Sextans~B,
respectively.  These M$_{\rm H_2}$ values correspond to upper limits
to the M$_{\rm H_2}$/M$_{\rm HI}$ ratios of 33\%, 39\%, and 13\%,
respectively.  Computing the star formation rate using the H$\alpha$
luminosity is notoriously unreliable at low metallicities
\citep{lee09}.  If instead we use the SFR$_{\rm FUV}^{Reg}$ values
given in Table \ref{galprops}, we compute M$_{\rm H_2}$ =
1.0$\times$10$^{7}$ M$_{\odot}$ and 2.3$\times$10$^{6}$ M$_{\odot}$
for Sextans~A and Sextans~B, respectively.  These values correspond to
M$_{\rm H_2}$/M$_{\rm HI}$ ratios of 114\% and 49\%.

In Table \ref{halcopreds} we compute the resulting CO luminosities
with our previous $\alpha_{\rm CO}$ assumptions.  If an $\alpha_{\rm
  CO}$ of 4.3 is appropriate, we would have detected this amount of
\Htwo\ in each of our galaxies with our observations.  In fact, we
would have detected Sextans~A using any of the $\alpha_{\rm CO}$
values with either SFR indicator.  Either the amount of CO present in
these galaxies is much lower than the upper limits calculated from the
SFRs or our assumptions of a constant SFR and 2 Gyr depletion
timescale do not accurately describe the local physics.  It is likely
the depletion timescale is much longer than 2 Gyr in low metallicity
environments. The L$_{\rm CO}$ values computed here are at least an
order of magnitude more than those predicted by the KMT+ model for
$\tau_{dep}$ = 2 Gyr.  The KMT+ model allows for the depletion
timescale to be as long as $\sim$100 Gyr in the \HI-dominated regions
similar to what we expect in our systems.  Increasing the
  depletion timescale will, likewise, increase the derived L$_{\rm
    CO}$ values.  The data seem to favor larger values of $\alpha_{\rm
    CO}$ but predicting an exact value requires more detailed
  understanding about the appropriate depletion timescales in each
  system.

\section{Conclusions}

We present CARMA CO (J = 1 $\rightarrow$ 0) observations of three
nearby, low-metallicity galaxies: Leo~P, Sextans~A, and Sextans~B.  We
do not detect any CO emission but derive very sensitive upper limits.
We use the KMT+ models presented in \citet{krum13} and some reasonable
physical assumptions to calculate a prediction for the estimated total
mass of \Htwo\ in the three galaxies, as well as the expected CO
luminosity.  We find that even under optimistic circumstances, CO will
be extremely difficult to observe in Leo~P with current technology.
On the other hand, CO luminosities as low as ~100 K km s$^{-1}$
pc$^{-2}$ can be detected with a modest amount of ALMA time, so
Sextans~A and Sextans~B should be observable with ALMA.  If future
observations do find CO emission in Leo~P, then this may imply that
the parameters in the KMT+ model and/or our assumptions about the
physical conditions in Leo~P are flawed in some way.

Even though CO (1-0) has been the main tracer of molecular material in high-metallicity
galaxies, other tracers of molecular gas need to be explored in low-metallicity systems.
We suggest several useful observations that can trace molecular material in these
systems.  
\begin{enumerate}
\item If the abundance of CO is low and the emission is optically
  thin, observations of CO (J~=~2~$\rightarrow$~1) may be a more
  sensitive probe of molecular gas in low-metallicity galaxies.  The
  excitation conditions of the 2-1 transition are not particularly
  stringent, and the Rayleigh-Jeans brightness temperature can be up
  to 4 times higher than for the 1-0 transition in warm gas in LTE for
  optically thin emission.

\item Continuum observations at sub-mm wavelengths (e.g., with ALMA
  Band 9) would allow for a derivation of the dust content.  It is
  interesting to note that the optical imaging and spectroscopy
  presented in \citet{rhod13}, \citet{mcqu13}, and \citet{skil13} each
  allow for the presence of a modest amount of differential extinction
  within Leo\,P.  If dust is detected in Leo\,P, then this can be used
  to infer the presence of molecular material.  We note that the
  observations of the Local Group galaxy WLM presented in
  \citet{jack04} made exactly such a prediction; the subsequent
  detections of CO in this system \citep{elme13} now represent the
  most metal-poor CO measurement to date.

\item Observations of warm \Htwo\ emission lines in the infrared would
  allow for a direct search for molecular material; while there would
  remain uncertainties about the excitation and temperature of the
  gas, such a detection of H$_{\rm 2}$ that is co-spatial with the
  \HI\ maximum and star formation would provide the possibility for
  cooler \Htwo\ as well.

\item Emission from the [C\,II] 158 $\mu$m line has been interpreted
  as a tracer of molecular gas in metal-poor environments
  \citep{madd97}.  Observations in this transition are possible with
  current instrumentation (SOFIA), and could confirm the theoretical
  predictions described in the models discussed above
  \citep[e.g.,][]{bola99}.
\end{enumerate}

\acknowledgements{We thank the anonymous referee for a prompt and 
  detailed report which significantly improved the clarity of the 
  manuscript.  S.R.W. is grateful to Lee Mundy for helpful
  conversations regarding the analysis in this work and Mark Krumholz
  for providing his gas modeling software. S.R.W.  would also like to
  thank Andreas Schruba for helping with the observing, setup scripts,
  and data reduction tips. The Undergraduate ALFALFA team is supported
  by NSF grants AST-0724918, AST-0725267, AST-0725380, AST- 0902211,
  and AST0903394. The ALFALFA work at Cornell is supported by NSF
  grants AST-0607007 and AST- 1107390 to RG and MPH and by grants from
  the Brinson Foundation. J.M.C. is supported by NSF grant AST-1211683
  and K.L.R. is supported by NSF Faculty Early Career Development
  (CAREER) award AST-0847109..

This research has made use of NASA's Astrophysics Data System
Bibliographic Services and the NASA/IPAC Extragalactic Database (NED),
which is operated by the Jet Propulsion Laboratory, California
Institute of Technology, under contract with the National Aeronautics
and Space Administration.

Support for CARMA construction was derived from the Gordon and Betty
Moore Foun- dation, the Kenneth T. and Eileen L. Norris Foundation,
the James S. McDonnell Foundation, the Associates of the California
Institute of Technology, the University of Chicago, the states of
California, Illinois, and Maryland, and the National Science
Foundation. Ongoing CARMA development and operations are supported by
the National Science Foundation under a cooperative agreement, and by
the CARMA partner universities.}

\clearpage
\begin{deluxetable}{lccc}  
\tabletypesize{\small}
\tablecaption{Properties of Galaxy Sample} 
\tablewidth{0pt}  
\tablehead{
\colhead{Parameter} &\colhead{Leo~P}  &\colhead{Sextans~A} &\colhead{Sextans~B}} 
\startdata
      
Right ascension (HH:MM:SS J2000) & 10:21:45   & 10:11:07  & 09:59:59 \\
Declination (DD:MM:SS J2000)     &+18:05:15   & -04:42:16 & +05:19:44 \\
Distance (Mpc)                   &1.62 $\pm$0.15\tablenotemark{a}  &1.38$\pm$0.05\tablenotemark{b} &1.39$\pm$0.04\tablenotemark{b} \\
M$_B$ (Mag.)                     &-8.97$\pm$0.10\tablenotemark{c} &-13.71$\pm$0.08\tablenotemark{b} &-13.88$\pm$0.06\tablenotemark{b} \\
Metallicity (12+log(O/H))        &7.17$\pm$0.04\tablenotemark{d} &7.54$\pm$0.06\tablenotemark{e} &7.53$\pm$0.05\tablenotemark{e} \\
M$_{\rm HI}^{\rm Tot}$ (10$^{6}$ M$_{\sun}$)\tablenotemark{f} & 0.81\tablenotemark{a} & 62.1\tablenotemark{h} & 41.5\tablenotemark{h}\\
\HI\ velocity range (\kms) & 250-294\tablenotemark{g} & 279-366\tablenotemark{h} & 259-346\tablenotemark{h}\\
L$_{\rm H\alpha}^{\rm Tot}$ (log erg s$^{-1}$)& 36.79\tablenotemark{c} & 38.66\tablenotemark{i} & 38.20\tablenotemark{i} \\
SFR$_{\rm H\alpha}^{\rm Tot}$ (10$^{-4}$ M$_{\sun}$ yr$^{-1}$)\tablenotemark{j} & 0.5 & 36.3 & 12.6 \\
SFR$_{\rm FUV}^{\rm Tot}$\tablenotemark{k} (10$^{-4}$ M$_{\sun}$ yr$^{-1}$)& \nodata & 120.2 & 51.3\\
\sidehead{Observed region properties:\tablenotemark{m}}
M$_{\rm HI}^{\rm Reg}$ (10$^{6}$ M$_{\sun}$) & 0.3 & 8.8 & 4.7 \\
L$_{\rm H\alpha}^{\rm Reg}$ (log erg s$^{-1}$)& 36.79 & 38.33 & 37.59 \\
SFR$_{\rm H\alpha}^{\rm Reg}$ (10$^{-4}$ M$_{\sun}$ yr$^{-1}$)\tablenotemark{j} & 0.5 & 17.0 & 3.1\\
SFR$_{\rm FUV}^{\rm Reg}$ (10$^{-4}$ M$_{\sun}$ yr$^{-1}$)\tablenotemark{m,n} & \nodata & 50.0 & 11.5
\enddata     
\label{galprops}
\begin{small}
\tablenotetext{a}{\citet{mcqu15}}
\tablenotetext{b}{\citet{dalc09}}
\tablenotetext{c}{\citet{rhod13}}
\tablenotetext{d}{\citet{skil13}}
\tablenotetext{e}{\citet{knia05}}
\tablenotetext{f}{Total galaxy \HI\ mass.}
\tablenotetext{g}{\citet{bern14}}
\tablenotetext{h}{\citet{ott12}}
\tablenotetext{i}{\citet{kenn08}}
\tablenotetext{j}{Estimated by converting the H$\alpha$ luminosity to a SFR via the relation in \citet{kenn98}.}
\tablenotetext{k}{\citet{lee09}}
\tablenotetext{m}{Values computed within half-power beam radius of our observed regions.}
\tablenotetext{n}{Assuming a SFR$_{\rm H\alpha}$/SFR$_{\rm FUV}$ ratio given in \citet{lee09} of 0.34 and 0.27 for
Sextans~A and Sextans~B, respectively.}
\end{small}
\end{deluxetable}   

\clearpage

\begin{deluxetable}{lccc}  
\tablecaption{Observed Galaxy Properties} 
\tablewidth{0pt}  
\tablehead{
\colhead{Parameter} &\colhead{Leo~P}  &\colhead{Sextans~A} &\colhead{Sextans~B}} 
\startdata
      
Velocity Resolution (\kms)     &18.4 &18.4 &18.4 \\
Beam size($\arcsec$)         & 3.49$\times$2.55 & 4.15$\times$2.52 & 3.38$\times$2.64\\
Linear resolution (pc)       & $\sim$27$\times$20 & $\sim$28$\times$17 & $\sim$23$\times$18\\
rms Noise (mJy beam$^{-1}$ channel$^{-1}$)     & 6.14 & 36.0 & 27.7 \\
4$\sigma$ S$_{\rm CO}$ Upper Limit (Jy beam$^{-1}$ km s$^{-1}$)     & 0.45 & 2.65 & 2.04 \\
L$_{\rm CO}$ Upper Limit (K km s$^{-1}$ pc$^2$)     & 2900 & 12400 & 9700 \\
\enddata     
\label{obsprops}
\end{deluxetable}   

\clearpage

\begin{table} 
\begin{center}
\caption{\Htwo\ Mass Upper Limits \label{h2masses}} 
\begin{tabular}{lcccccc}
\hline\hline
& \multicolumn{2}{c}{$\alpha_{\rm CO}$ = 4.3$^{a}$} & \multicolumn{2}{c}{$\alpha_{\rm CO}$ =
70$^{b}$} & \multicolumn{2}{c}{$\alpha_{\rm CO}$ = 285$^{c}$} \\
\cmidrule(r{0.25em}){2-3} \cmidrule(r{0.25em}){4-5} \cmidrule(r{0.25em}){6-7}
Galaxy & M$_{\rm H_{2}}$ (M$_{\odot}$) & M$_{\rm H_{2}}$/M$_{\rm HI}$ & M$_{\rm H_{2}}$ (M$_{\odot}$) & M$_{\rm H_{2}}$/M$_{\rm HI}$ & M$_{\rm H_{2}}$ (M$_{\odot}$) & M$_{\rm H_{2}}$/M$_{\rm HI}$\\
\hline
Leo~P     & 1.2$\times$10$^{4}$ & 0.015 & 2.0$\times$10$^{5}$ & 0.25 & 8.3$\times$10$^{5}$ & 1.0 \\      
Sextans~A & 5.3$\times$10$^{4}$ & 0.006 & 8.7$\times$10$^{5}$ & 0.10 & 3.5$\times$10$^{6}$ & 0.40 \\
Sextans~B & 4.2$\times$10$^{4}$ & 0.009 & 6.8$\times$10$^{5}$ & 0.14 & 2.8$\times$10$^{6}$ & 0.60 \\
\hline
\end{tabular}
\end{center}

$^{a}$Galactic value; 
$^{b}$$\alpha_{\rm CO}$ = SMC value \citep{lero09}; 
$^{c}$$\alpha_{\rm CO}$ = Value taken from the 3\% Solar metallicity models of
\citet{glov12b}.
\end{table}   

\clearpage

\begin{deluxetable}{lcccccc}  
\tablecaption{\Htwo\ Mass Predictions from \citet{krum13}} 
\tablewidth{0pt}  
\tablehead{\colhead{Galaxy} &\colhead{$\rho_{sd}$}  &\colhead{M$_{\rm H_{2}}$} &\colhead{$\rho_{sd}$} &\colhead{M$_{\rm H_{2}}$} &\colhead{$\rho_{sd}$} &\colhead{M$_{\rm H_{2}}$}\\
                            &\colhead{(M$_{\odot}$/pc$^3$)}  &\colhead{(M$_{\odot}$)} &\colhead{(M$_{\odot}$/pc$^3$)} &\colhead{(M$_{\odot}$)} &\colhead{(M$_{\odot}$/pc$^3$)} &\colhead{(M$_{\odot}$)}}
\startdata

Leo~P     & 0.001 & 34 & 0.01 & 87 & 0.1 & 257 \\      
Sextans~A & 0.001 & 1.1$\times$10$^5$ & 0.01 & 1.3$\times$10$^5$ & 0.1 & 2.2$\times$10$^5$ \\
Sextans~B & 0.001 & 1.4$\times$10$^4$ & 0.01 & 2.2$\times$10$^4$ & 0.1 & 5.1$\times$10$^4$ \\
\enddata     
\label{krumh2masses}
Note --- These values are derived from the models of \citet{krum13} which require inputs of $\Sigma_{gas}$, $\rho_{sd}$, Z, and cf (see
\S\ref{h2masses} for details).  We have assumed values for Z listed in Table \ref{galprops} and a clumping factor cf = 1.
\end{deluxetable}   

\clearpage

\begin{deluxetable}{lccccc}
%\rotate
\tablecaption{L$_{\rm CO}$ Predictions from \citet{krum13}}
%\tabletypesize{\scriptsize}
\tablewidth{0pt}  
\tablehead{\colhead{Galaxy} & &\colhead{$\alpha_{\rm CO}$ = 4.3} &\colhead{$\alpha_{\rm CO}$ = 70}  &\colhead{$\alpha_{\rm CO}$ =
285} & \colhead{Observed$^{a}$}}
\startdata

Leo~P     & L$_{\rm CO}$ = & 20$^{+39}_{-12}$                    & 1.2$^{+1.5}_{-0.7}$                 & 0.30$^{+0.60}_{-0.18}$ &
2900 \\      
Sextans~A & L$_{\rm CO}$ = & 3.0$^{+2.1}_{-0.5}$$\times$10$^{4}$ & 1860$^{+1280}_{-290}$ & 460$^{+310}_{-70}$ & 12400\\
Sextans~B & L$_{\rm CO}$ = & 5100$^{+6900}_{-1900}$ & 310$^{+420}_{-110}$  & 77$^{+100}_{-28}$ & 9700\\
\enddata     
\label{lcopreds}
\tablecomments{Based upon \Htwo\ mass estimates from the $\rho_{sd}$ = 0.01 models with errors reflecting the 
spread in values from the $\rho_{sd}$ = 0.001 and 0.1 models in Table \ref{krumh2masses}.\\
$^{a}$Observed values from Table \ref{obsprops}}
\end{deluxetable}   

\clearpage

\begin{deluxetable}{lcccccc}
%\rotate
\tablecaption{L$_{\rm CO}$ Predictions from SFR indicators}
%\tabletypesize{\scriptsize}
\tablewidth{0pt}  
\tablehead{\colhead{Galaxy} &\colhead{log(M$_{\rm H_2}$) (M$_{\odot}$)} & &\colhead{$\alpha_{\rm CO}$ = 4.3} &\colhead{$\alpha_{\rm
CO}$ = 70}  &\colhead{$\alpha_{\rm CO}$ = 285} &\colhead{Observed$^{a}$}}
\startdata
\sidehead{Using SFR$_{\rm H\alpha}^{\rm Reg}$}
Leo~P     & 5.0  & L$_{\rm CO}$ = & 2.3$\times$10$^{4}$ & 1.4$\times$10$^{3}$ & 350 & 2900\\      
Sextans~A & 6.53 & L$_{\rm CO}$ = & 7.9$\times$10$^{5}$ & 4.9$\times$10$^{4}$ & 1.2$\times$10$^{4}$ & 12400\\
Sextans~B & 5.79 & L$_{\rm CO}$ = & 1.4$\times$10$^{5}$ & 8.9$\times$10$^{3}$ & 2.1$\times$10$^{3}$ & 9700 \\
\sidehead{Using SFR$_{\rm FUV}^{\rm Reg}$}
Sextans~A & 7.0  & L$_{\rm CO}$ = & 2.3$\times$10$^{6}$ & 1.4$\times$10$^{5}$ & 3.5$\times$10$^{4}$ \\
Sextans~B & 6.36 & L$_{\rm CO}$ = & 5.3$\times$10$^{5}$ & 3.3$\times$10$^{4}$ & 8.0$\times$10$^{3}$ \\
\enddata     
\label{halcopreds}
\tablecomments{Based upon \Htwo\ mass upper limits computed by multiplying the region SFR values from
Table \ref{galprops} by $\tau_{dep}$ = 2 Gyr.\\
$^{a}$Observed values from Table \ref{obsprops}}
\end{deluxetable}   

\clearpage

\clearpage

\clearpage

\end{document}